\documentclass{article}
\usepackage{spconf,amsmath,graphicx}
\usepackage{hyperref}
\usepackage[T1]{fontenc}
\usepackage[utf8]{inputenc}
\usepackage{amsmath} 
\usepackage{amsfonts} 
\usepackage{amssymb} 
\usepackage{amsthm}
\usepackage{color}
\usepackage{url}
\usepackage{subcaption}
\usepackage{float}
\usepackage[linesnumbered,ruled,vlined]{algorithm2e}
\usepackage{cleveref}
\usepackage{verbatim}
\usepackage{multirow}
\usepackage{graphicx}
\usepackage{comment}
\SetKwInput{KwInput}{Input}                
\SetKwInput{KwOutput}{Output}              
\SetKwInput{Kwinitialize}{Initialization}              

\renewcommand{\vec}[1]{{\bf{#1}}} 
\newcommand{\vecgreek}[1]{{\boldsymbol{#1}}} 
\newcommand{\tran}{^{\mbox{\scriptsize T}}}
\newcommand{\herm}{^{\mbox{\scriptsize H}}}

\newcommand{\fro}[1]{\| #1\|_{\mathrm{F}}^2}
\newcommand{\norm}[1]{\| #1\|}

\newcommand{\gammab}{\vecgreek{\gamma}}
\newcommand{\Thetab}{\vecgreek{\Theta}}

\newcommand{\Psib}{\vecgreek{\Psi}}

\newcommand{\Sigmab}{\vecgreek{\Sigma}}

\newcommand{\Lamdbab}{\vecgreek{\Lambda}}
\newcommand{\mybibliography}{\bibliography{conf_short,jour_short,refs}}

\title{Joint Estimation of Clustered User Activity and Correlated Channels with Unknown Covariance in mMTC}

\name{Hamza Djelouat, Markus Leinonen, and Markku Juntti \thanks{This work has been financially supported in part by the Academy of Finland (ROHM project, grant 319485) and (6G Flagship program, grant 346208). The work of M. Leinonen has also been financially supported in part the Academy of Finland (grant 340171).}}
\address{Centre for Wireless Communications -- Radio Technologies, FI-90014, University of Oulu, Finland}

\begin{document}
\ninept
\maketitle
\begin{abstract}
This paper considers joint user identification and channel estimation (JUICE) in grant-free access with a \emph{clustered} user activity pattern. In particular, we address the JUICE in massive machine-type communications (mMTC) network under correlated Rayleigh fading channels with unknown channel covariance matrices. We formulate the JUICE problem as a maximum \emph{a posteriori} probability (MAP) problem with properly chosen priors to incorporate the partial knowledge of the UEs' clustered activity and the unknown covariance matrices. We derive a computationally-efficient algorithm based on alternating direction method of multipliers (ADMM) to solve the MAP problem iteratively via a sequence of closed-form updates. Numerical results highlight the significant improvements brought by the proposed approach in terms of channel estimation and activity detection performances for clustered user activity patterns.

\end{abstract}
\section{Introduction}
Sparse signal recovery techniques have  widely been  applied in the design of massive machine-type communications (mMTC) solutions with grant-free access protocols. A major challenge in grant-free access is the joint user identification and channel estimation (JUICE). Subsequently, owing to the sporadic nature of activity pattern of the mMTC devices, namely user equipments (UEs), the JUICE has been widely addressed as a sparse recovery problem and solved using various algorithms such as approximate message passing (AMP), sparse Bayesian learning (SBL), and mixed-norm minimization.

The vast majority of the works address the JUICE in an uncorrelated channel model \cite{chen2018sparse,liu2018massive,liu2018massiveII,senel2018grant,ke2020compressive,zhang2017novel}. However, this assumption is not practical as the multiple-input multiple-output (MIMO)  channels are usually spatially correlated \cite{sanguinetti2019towards}. Recently, few works have addressed the JUICE in the correlated MIMO channels, thereby better reflecting the reality, using AMP in \cite{cheng2020orthogonal,9851841}  and using mixed-norm minimization  in \cite{Djelouat-Leinonen-Juntti-21-icassp,djelouat2021spatial,Djelouat2020Joint}. Nonetheless, these works assume that channel distribution information (CDI) for all the UEs is available at the base station (BS), which can be challenging to acquire in realistic environments. Furthermore, the prior work on the JUICE  considers mMTC networks with a random  UE activity pattern. This could model, e.g., a scenario where UEs monitor independent random processes and thus activate randomly based on certain application criteria. 

This paper makes the following two distinctions from the prior works: First, we address the JUICE in  spatially correlated MIMO channels with \emph{no prior} knowledge of the exact CDI. Second,  we consider an mMTC network where the UEs activity has a specific structure, namely, a \textit{clustered} form.  For instance, this models a network where the UEs form clusters based on their geographical locations and each cluster is associated with a monitoring task. Here, an event could trigger a small subset of UEs belonging to a cluster to activate concurrently, leading to clustered UE activity system-wise.

This paper adopts a Bayesian approach and formulates the JUICE as a maximum \emph{a posteriori} probability (MAP) estimation problem  that takes into consideration the different  specific structures of the system model.  In particular, to accommodate the clustered activity of the UEs, we propose a log-sum based prior function composed of two different components   to capture both the overall sparsity and the clustered sparsity  of the signal model. Furthermore, in order to estimate the CDI as well, we utilize a  prior function that follows a Wishart distribution. Subsequently, we derive a computationally-efficient iterative solution based on alternating direction method of multipliers (ADMM) that solves an approximate version of the MAP problem as a sequence of closed-form updates.

\section{System Model}
We consider a single-cell uplink network consisting of $N$ UEs served by a single BS equipped with a uniform linear array (ULA) of $M$ antennas. The UEs are geographically distributed so that they form $C$ \textit{clusters}, where each cluster contains $L$ UEs\footnote{For the simplicity of presentation, yet without loss of generality, we assume that all clusters contain the same number of UEs.}, such that ${N=LC}$. 
A cluster containing a subset of UE indices is denoted by $\mathcal{C}_{l}\subseteq\{1,2,\ldots,N\}$.


The propagation channel between the $i$th UE and the BS, denoted by $\vec{h}_i$, follows a local scattering model \cite{massivemimobook}. Thus, each $\vec{h}_i$ is modelled as a zero-mean complex Gaussian random variable, i.e., ${\vec{h}_i \sim  \mathcal{CN}(0,\Sigmab_i^{-1})}$,  where $\Sigmab_i \in  \mathbb{C}^{M \times M}$ denotes the channel precision matrix, i.e., the inverse covariance matrix  ${\Sigmab_i^{-1}=\mathbb{E}[\vec{h}_i\vec{h}_i\herm]}$. We consider UEs with low mobility, which is justified in 
mMTC, e.g., in applications like industrial monitoring. Hence, we adopt the common assumption that the channels are  wide-sense stationary. Thus,  the  set of channel precision matrices ${\Sigmab=\{\Sigmab_i\}_{i=1}^N}$  vary in a slower timescale  compared to the channel realizations \cite{sanguinetti2019towards}. 


Since the UE activation is sporadic in mMTC,
only a small number of the $N$ UEs are active at each coherence time $T_\mathrm{c}$. To this end, the UE activity indicator vector ${\gammab=[\gamma_1,\ldots,\gamma_N]\tran}$ is given as
\begin{equation}\footnotesize\label{eq_UE_sparsity}
    \gamma_i=\begin{cases}
1, & i\mathrm{th} \text{  UE is active}\\ 
0 , & \text{otherwise},
\end{cases}\qquad \forall i=1,\ldots,N.
\end{equation}
Contrary to the majority of the literature on mMTC that consider random UE activation, we consider that the UEs activate in a specific \textit{clustered} manner. More precisely, the active UEs  belong to a small (random) number of active clusters, where an active cluster refers to any cluster with at least one active UE, while containing at most ${L_\mathrm{c}\leq L}$ active UEs. Such cluster sparsity imposes a special structure on $\gammab=[\underbrace{\gamma_1,\ldots,\gamma_L}_{\footnotesize\mbox{Cluster~1} },\underbrace{\gamma_{L+1},\ldots,\gamma_{2L}}_{\footnotesize\mbox{Cluster~2}},\ldots,\underbrace{\gammab_{N-L+1},\ldots,\gammab_{N}}_{\footnotesize\mbox{Cluster~}C}]$ where the elements belonging to a given cluster ($\gamma_i$, $i\in\mathcal{C}_l$) are assumed to be correlated. The correlation in the UE activity is imposed by the priors presented in Section \ref{sec:p(gamma)}.





For data transmission, each UE $ i\in \mathcal{N}$ is assigned a unique unit-norm pilot sequence $\vecgreek{\phi}_i \in \mathbb{C}^{\tau_{\mathrm{p}}}$, as well as a transmit  power $p_i$ that is inversely proportional to its average channel gain \cite{senel2018grant,bjornson2016massive}. 
Accordingly, the received pilot signal $\vec{Y} \in \mathbb{C}^{\tau_{\mathrm{p}}\times M}$ is given by
\begin{equation}\footnotesize
\label{eq::Y}
\vec{Y}=\sum_{i=1}^{N}\gamma_i \sqrt{p_i}   \vecgreek{\phi}_i\vec{h}_i\tran+\vec{W}=\vec{\Phi} \vec{X}\tran+ \vec{W},
\end{equation}
where  ${\vec{X}=[\vec{x}_1,\ldots,\vec{x}_{N}] \in \mathbb{C}^{M\times N}}$ denotes the effective channel matrix with ${\vec{x}_i=\gamma_i \sqrt{p_i} \vec{h}_i}$, $\vec{\Phi}=[\vecgreek{\phi}_1,\ldots,\vecgreek{\phi}_N] \in \mathbb{C}^{\tau_{\mathrm{p}}\times N}$ is the  pilot book, and $\vec{W}\sim \mathcal{CN}(0,\,\sigma^{2}\vec{I}_M) \in\mathbb{C}^{\tau_{\mathrm{p}}\times M}$ represents additive white Gaussian noise.

\section{JUICE with Clustered Activity and Partially Known CDI}
The signal model in \eqref{eq::Y} represents  a linear measurement where the (unknown) effective channel matrix $\vec{X}\tran$  exhibits row sparsity.  Thus, detecting the active UEs and estimating their channel gives rise to a sparse recovery problem from a multiple measurement vector (MMV) setup, which can be solved via several CS algorithms. For instance, the Bayesian AMP algorithm \cite{kim2011belief} can be utilized if 
the statistics of a sparse signal are known a priori. However, when such  information is not available,  mixed-norm minimization \cite{steffens2018compact} using deterministic sparsity regularization is often utilized. Clearly, the additional prior information in AMP leads to superior performance in general. However, attaining prior CDI  is a nontrivial task and requires network resources to perform a training phase.

In this paper, we aim to tackle the aforementioned issue related to the acquisition of CDI by adopting a Bayesian framework which aims 
at not only detecting the active UEs and estimate their channels, but also estimating their covariance matrices.
More precisely, by formulating the JUICE as an MAP estimation problem and using the adequate prior functions,  we develop a computationally-efficient algorithm that solves a relaxed version of the MAP problem iteratively via a set of closed-form update rules, as we show next.

    

The MAP estimates $\{\hat{\vec{X}},\hat{\gamma},\hat{\Sigmab}\}$ with respect to the posterior density
given the measurement matrix $\vec{Y}$ is given by
\begin{equation}\footnotesize
\begin{array}{ll}\label{eq:map_x_gamma}
\{\hat{\vec{X}},\hat{\gamma},\hat{\Sigmab}\}&\hspace{-3mm}=\underset{\vec{X},\gamma,\Sigmab}{\max}~\displaystyle p(\vec{X},\gamma,\Sigmab|\vec{Y})\\
&\hspace{-3mm}=\underset{\vec{X},\gamma,\Sigmab}{\max}~\displaystyle\frac{p(\gamma,\Sigmab)p(\vec{X}|\gamma,\Sigmab)p(\vec{Y}|\vec{X},\gamma,\Sigmab)}{p(\vec{Y})}\\&\hspace{-3mm}\overset{(a)}{=}\underset{\vec{X},\gamma,\Sigmab}{\max}~\displaystyle{p(\gamma,\Sigmab)p(\vec{X}|\gamma,\Sigmab)p(\vec{Y}|\vec{X})}\\
&\hspace{-3mm}\overset{}{=}\underset{\vec{X},\gamma,\Sigmab}{\min}~\displaystyle{-\log\,p(\vec{Y}|\vec{X})}-\log\,p(\vec{X}|\gamma,\Sigmab)-\log\,p(\gamma,\Sigmab)\\
&\hspace{-3mm}\overset{(b)}{=}\underset{\vec{X},\gamma,\Sigmab}{\min}~\displaystyle\frac{1}{\sigma^2}\|\vec{Y}-\vec{\Phi}\vec{X}\tran\|_{\mathrm{F}}^{2}-\log\,p(\vec{X}|\gamma,\Sigmab)\\&-\log\,p(\Sigmab)-\log\,p(\gamma) \end{array}
\end{equation}
where $(a)$ follows from the Markov chain ${\{\gammab,\vec{\Sigmab}\}\rightarrow\vec{X}\rightarrow\vec{Y}}$ and  because $p(\vec{Y})$ does not affect the maximization and $(b)$ follows from the additive Gaussian noise model in \eqref{eq::Y} and independence of UE activity and channels. 
Next,  we elaborate  the definition of  the conditional PDF $p(\vec{X}|\gammab)$ and  the choice of the priors $p(\Sigmab)$  and $p(\gammab)$.

\subsection{Conditional PDF $ p(\vec{X}|\gammab,\Sigmab)$}
Since the user activity is controlled by $\gammab$, the conditional PDF $p(\vec{x}_i|\gamma_i,\Sigmab_i)$ is defined as follows:
\begin{equation}\footnotesize
       p(\vec{x}_i|\gamma_i,\Sigmab_i)= \mathcal{CN}(\vec{x}_i;\vec{0},p_i\Sigmab_i^{-1})^{\gamma_i}\mathbb{I}(\vec{x}_i=\vec{0})^{1-\gamma_i},
\label{pdf_x}
\end{equation}
where $\mathbb{I}(a)$ is an indicator function that takes the value 1 if $a\neq0$, and 0 otherwise.  $p(\vec{x}_i|\gamma_i,\Sigmab_i)$ implies that
conditioned on $\gamma_i=0$, $\vec{x}_i$ is equal to 0 with probability one, and conditioned on  ${\gamma_i=1}$, $\vec{x}_i$  follows a complex Gaussian distribution with zero mean  and precision matrix $\Sigmab_i^{-1}$. Furthermore, since $\vec{x}_i$ depends only on $\gamma_i$ and $\Sigmab_i$,  ${-\log p(\vec{X}|\gammab,\Sigmab)=\sum_{i=1}^N-\log p(\vec{x}_i|\gamma_i,\Sigmab_i)}$ is given as
\begin{equation}\label{PDF_X}\footnotesize
     -\log p(\vec{X}|\gammab,\Sigmab) 
     \propto  \displaystyle\sum_{i=1}^N - \gamma_i p_i^M\log  |\Sigmab_i|+  \frac{\gamma_i}{p_i} \vec{x}_i\herm \Sigmab_i\vec{x}_i.
\end{equation}

\subsection{Prior on Precision Matrix}
A common and physically grounded prior for an  unknown precision matrix $\Sigmab_i$ of the Gaussian random variable $\vec{x}_i$  is given by the  Wishart distribution \cite{bishop2006pattern}, defined as
\begin{equation}\label{p_Sigmab}\footnotesize
    p(\Sigmab_i)\sim \mathcal{W}(\Sigmab_i|\vec{B}_i,v)= f(\vec{B}_i,v)  |\Sigmab_i|^d \exp\left(-\text{Tr}(\vec{B}_i^{-1} \Sigmab_i)\right)
\end{equation}
where $f(\vec{B}_i,v)$ is a normalization constant given by \cite[eq (B.79)]{bishop2006pattern}, ${d=v-M+1>0}$ where $v$  controls the degrees of freedom of the distribution, and ${\vec{B}_i \in \mathbb{C}^{M \times M}}$ is a symmetric, positive definite  matrix that represents the prior guess for the precision matrix $\Sigmab_i$. Subsequently, we write 
\begin{equation}\footnotesize\label{logWishart}
   -\log p(\Sigmab)  = \sum_{i=1}^N-\log p(\Sigmab_i) \propto \sum_{i=1}^N-d \log |\Sigmab_i|+ \text{Tr}(\vec{B}_i^{-1}\Sigmab_i), 
\end{equation}
where $-\log(f(\vec{B},v))$ was dropped as it does not depend on $\Sigmab_i$.

\subsection{Sparsity-promoting Prior $p(\gammab)$}\label{sec:p(gamma)}
For a sparse signal recovery problem, utilizing prior functions that promote sparsity while incorporating the possible specific structures of the sparse signal is the key to achieve accurate solutions. We discuss two different prior functions that capture the different structural features of our underlying sparse signal below. 

\textit{1) Separable prior:}
Since the UEs have a sparse activity pattern, the optimal sparsity-inducing prior is the $\ell_0$-norm penalty on $\gammab$, i.e., $\sum_{i=1}^{N}\mathbb{I}(\gamma_i)$. However, the $\ell_0$-norm penalty is intractable for large $N$. Thus, many surrogate functions have been proposed to relax it, for instance,  \emph{log-sum} penalty ${\sum_{i=1}^{N}\log(\gamma_i+\epsilon)}$, which is used herein. This resembles most closely the $\ell_0$-norm penalty when ${\epsilon_{0} \rightarrow 0}$. Subsequently, we define a sparsity prior function as 
\begin{equation}\footnotesize\label{eq:T1}
J_{\mathrm{s}}(\gammab)\propto \sum_{i=1}^{N}\log(\gamma_i+\epsilon_0).  
\end{equation}

\textit{2) Cluster-sparsity-promoting  prior:}
Although the prior \eqref{eq:T1} is an appropriate choice as it: 1) promotes sparsity, 2) is separable across the UEs, it ignores the considered clustered structure of the UEs' activity pattern.
To account for this correlated activity, we propose a cluster-sparsity-promoting prior that captures the  relations between the UE activity indicators belonging to the same cluster, i.e., $\gamma_i$, ${i\in \mathcal{C}_l}$. More precisely, we propose the following prior function:
\begin{equation}\footnotesize\label{eq:T2}
J_{\mathrm{c}}(\gammab)\propto \sum_{l=1}^ {C}\log\Big(\sum_{i \in \mathcal{C}_l}\gamma_i+\epsilon_0\Big). 
\end{equation}

While $J_{\mathrm{s}}(\gammab)$ is blind to the cluster sparsity structure, $J_{\mathrm{c}}(\gammab)$ promotes quite stringently solutions that have clustered sparsity as it has the tendency to enforce all UEs within each cluster to be detected active even if only one UE is active, being thereby susceptible to high false alarm error rate. Therefore, $J_{\mathrm{c}}(\gammab)$ would face  robustness issues in the instances where the UEs activity pattern does not exhibit a clustered structure.



\section{Proposed Solution Via ADMM}
For deriving the proposed solution, we make two technical choices:

    1) 
    Inspired by the argument raised in \cite{zhang2011sparse}, aiming to estimate  all $N$ precision matrices $\Sigmab_i$ with the available data at the BS  may lead to overfitting. Thus, we restrict the estimation of the $N$ precision matrices  $\Sigmab_i$ to $L$ precision matrices $\Sigmab_l$. 
    More precisely, we assume that all propagation channels of the UEs within a single cluster share the same CDI, i.e., $\vec{h}_i \sim \mathcal{CN}(\vec{0},\Sigmab_l^{-1})$, ${\forall i\in \mathcal{C}_l}$, ${\forall l=1,\ldots,C}$.
    
    2) 
    The binary nature of $\gammab$ renders the MAP estimation problem intractable for large $N$. To overcome this challenge, we note that finding the index set $\{i \mid \gamma_i \neq0,\;i\in\mathcal{N}\}$ is equivalent to finding the index set $\{i \mid \|\vec{x}_i\| >0,\;i\in\mathcal{N}\}$. Thus, we can eliminate the variable  $\gammab$ from the optimization problem by approximating each $\gamma_i$ by $\|\vec{x}_i\|$ and by relaxing $p(\gammab)$ by an equivalent prior function $p(\vec{X})$ that depends on  $\|\vec{x}_i\|$, $\forall i \in \mathcal{N}$,  as we will show in the next section.




By using the aforementioned arguments and   substituting \eqref{PDF_X} and \eqref{logWishart} into \eqref{eq:map_x_gamma}, the MAP estimation problem can be rewritten as
\begin{equation}\label{map_S}\footnotesize
\begin{array}{ll}
       \!\!\{\hat{\vec{X}},\hat{\Sigmab}\}\!=\!\underset{\vec{X},\Sigmab}{\min}~\displaystyle\frac{1}{2}\|\vec{Y}-\vec{\Phi}\vec{X}\tran\|_{\mathrm{F}}^{2} -\beta_1\log p(\vec{X})+\beta_2\sum_{l=1}^C\sum_{i \in \mathcal{C}_l}\frac{\vec{x}_{i}\herm\Sigmab_l\vec{x}_{i}}{p_i} \\\displaystyle\!\!-\beta_2\sum_{l=1}^{C}  \log |\Sigmab_l| \sum_{i \in \mathcal{C}_l}p_i^M \|\vec{x}_i\| \!-\! \beta_3 L\sum_{l=1}^{C} \big( d \log |\Sigmab_l|\displaystyle \!+\!  \mathrm{tr}(\vec{B}_l^{-1}\Sigmab_l)\big),
\end{array}
\end{equation}
where regularization weights $\beta_1$, $\beta_2$,  and $\beta_3$ control the emphasis on the priors with respect to the measurement fidelity term. 





In the following, we propose an iterative solution by alternating $p(\vec{X})$ over $J_{\mathrm{c}}(\cdot)$ and $J_{\mathrm{s}}(\cdot)$. The central idea is to develop an iterative two-level algorithm, whose outer loop aims at detecting the active clusters, and the inner loop aims at detecting the active UEs in each of the estimated active cluster. More precisely, in the outer loop, the algorithm enforces the detection of active clusters via the cluster-sparsity-promoting prior $J_{\mathrm{c}}$ in \eqref{eq:T2}. Subsequently, the algorithm runs an inner loop over the just-estimated active clusters to detect the individual active UEs belonging to them by using the sparsity-promoting prior $J_{\mathrm{s}}$ in \eqref{eq:T1}. The algorithm details are presented next.



\subsection{Outer Loop}
The outer loop aims to detect the set of the active clusters, hence, we enforce $p(\vec{X})$ to promote the cluster-sparsity by $- \log p(\vec{X})=\sum_{l=1}^ {C}\log(\sum_{i \in \mathcal{C}_l}\|\vec{x}_i\|+\epsilon_0)$. Since $-\log p(\vec{X})$ is concave, we apply a majorization-minimization (MM) approximation to linearize $- \log p(\vec{X})\approx \sum_{i=1}^N q_i^{(k_{\mathrm{c}})}\|\vec{x}_i\|$, 
where $q_i^{(k_{\mathrm{c}})}=\big(\sum_{i \in \mathcal{C}_l}\|\vec{x}_i^{(k_{\mathrm{c}})}\|+\epsilon_0\big)^{-1}$ and  $k_{\mathrm{c}}$ is the MM iteration index  for the outer loop. 
Thus,  the relaxed version of the  problem \eqref{map_S} can be solved iteratively as  
\begin{equation}\footnotesize
\begin{array}{ll}
      &\{\hat{\vec{X}}^{(k_{\mathrm{c}}+1)},\Sigmab^{(k_{\mathrm{c}}+1)}\} =\displaystyle\underset{\vec{X},\Sigmab}{\min}~\displaystyle\frac{1}{2}\|\vec{Y}-\vec{\Phi}\vec{X}\tran\|_{\mathrm{F}}^{2}+\beta_1 \sum_{i=1}^N q_i^{(k_{\mathrm{c}})} \|\vec{x}_i\| \\&\displaystyle+\beta_2\sum_{l=1}^C\sum_{i \in \mathcal{C}_l}\vec{x}_{i}\herm\Sigmab_l\vec{x}_{i}-\sum_{l=1}^{C} \mu^{(k_{\mathrm{c}})} \log |\Sigmab_l| +\beta_3 L\sum_{l=1}^{C} \mathrm{tr}(\vec{B}_l^{-1}\Sigmab_l),
\end{array}\label{map_K_c}
\end{equation}
 where ${\scriptsize \mu^{(k_{\mathrm{c}})}=\big(\beta_2  \displaystyle\sum_{i \in \mathcal{C}_l} p_i^M q_i^{(k_{\mathrm{c}})}\|\vec{x}_i^{(k_{\mathrm{c}})}\| + \beta_3 L d\big)}$.
 
We develop a computationally efficient ADMM solution for \eqref{map_K_c} through a set of sequential update rules, each computed in closed-form. We introduce two splitting variables ${\vec{Z},\vec{V}\in \mathbb{C}^{M\times N}}$ and the Lagrange dual variable matrices $\Lamdbab_v,\Lamdbab_z$ and define the set of variables to be estimated as $\Thetab=\{\vec{X},\Sigmab,\vec{Z},\vec{V},\Lamdbab_{\mathrm{z}},\Lamdbab_{\mathrm{v}}\}$. Subsequently, we write the augmented Lagrangian as
\begin{equation}\label{eq:lagrange}\footnotesize
  \begin{array}{ll}
       &\mathcal{L}(\Thetab)=\displaystyle\frac{1}{2}\|\vec{Y}-\vec{\Phi}\vec{Z}\tran\|_{\mathrm{F}}^{2} +\beta_1 \sum_{i=1}^N q_i^{(k_{\mathrm{c}})} \|\vec{x}_i\|+\beta_2\sum_{i=1}^{N}\vec{v}_{i}\herm\Sigmab_i\vec{v}_{i}\\\displaystyle &-\displaystyle\sum_{l=1}^{C}\big( \mu^{(k_{\mathrm{c}})} \log |\Sigmab_l|  +\beta_3 L \mathrm{tr}(\vec{B}_l^{-1}\Sigmab_l)\big)+\frac{\rho}{2}\|\vec{X}-\vec{V}+\frac{1}{\rho}\vecgreek{\Lambda}_{\mathrm{v}}\|_{\mathrm{F}}^2\\&+\displaystyle\frac{\rho}{2}\|\vec{X}-\vec{Z}+\displaystyle\frac{1}{\rho}\vecgreek{\Lambda}_{\mathrm{z}}\|_{\mathrm{F}}^2 -\displaystyle\frac{\fro{\vecgreek{\Lambda}_{\mathrm{z}}}}{2\rho}
     -\displaystyle\frac{\fro{\vecgreek{\Lambda}_{\mathrm{v}}}}{2\rho}.
\end{array}  
\end{equation}
ADMM solves to the optimization problem \eqref{map_K_c}  by minimizing the augmented Lagrangian  $\mathcal{L}(\Thetab)$ in \eqref{eq:lagrange}  over the primal variables $(\vec{Z},\vec{V},\vec{X},\Sigmab)$, followed by updating the dual variables $(\vecgreek{\Lambda}_{\mathrm{z}},\vecgreek{\Lambda}_{\mathrm{v}})$ \cite{boyd2011distributed}. Primal variable update is given by 
\begin{align}\label{eq::z(k_c+1)_sec2} \footnotesize
   \hspace{-0.21cm}  \vec{Z}^{(k_{\mathrm{c}}+1)}=
   \displaystyle\min_{\vec{Z}}\frac{1}{2}\| \vec{\Phi}\vec{Z}\tran-\vec{Y}\|_{\mathrm{F}}^2+ \frac{\rho}{2} \Vert\vec{X}^{(k_{\mathrm{c}})} -\vec{Z} +\frac{1}{\rho}\vecgreek{\Lambda}_{\mathrm{z}}^{(k_{\mathrm{c}})} \|_{\mathrm{F}}^2
\end{align}\vspace{-.7cm}
\begin{align}
\label{eq::v(k_c+1)_sec2}\footnotesize
\hspace{-0.16cm}\vec{V}^{(k_{\mathrm{c}}+1)}=\displaystyle\min_{\vec{V}}\beta_2  \sum_{i=1}^{N}  \vec{v}_i\herm\Sigmab_l^{(k_{\mathrm{c}})}\vec{v}_i  +\frac{\rho}{2} \Vert   \vec{X}^{(k_{\mathrm{c}})} - \vec{V} +\frac{\vecgreek{\Lambda}_{\mathrm{v}}^{(k_{\mathrm{c}})}}{\rho} \|_{\mathrm{F}}^2
\end{align}\vspace{-.8cm}
\begin{align}\label{eq::x(k_c+1)_sec} \footnotesize
\hspace{-1.5cm}\vec{X}^{(k_{\mathrm{c}}+1)}=\displaystyle\min_{\vec{X}}  \sum_{i=1}^{N} \alpha_i^{(k_{\mathrm{c}})} \Vert  \vec{x}_i\|+ \footnotesize\frac{\rho}{2} \|\vec{X}-\vec{C}^{(k_{\mathrm{c}}+1)} \|_{\mathrm{F}}^2
\end{align}\vspace{-.5cm}
\begin{align}\label{eq::sigma(k_c+1)_sec}
\begin{array}{ll}
\hspace{-.4cm}\Sigmab_l^{(k_{\mathrm{c}}+1)}\!\!=\!\displaystyle\min_{\Sigmab_l}  \beta_2\sum_{i \in \mathcal{C}_l}\!\vec{v}_{i}^{(k_{\mathrm{c}}+1)\herm}\!\Sigmab_l\vec{v}_{i}^{(k_{\mathrm{c}}+1)}\!-\!\mu^{(k_{\mathrm{c}}+1)}\! \log |\Sigmab_l| \\\,\,\,\,\qquad+\beta_3 L \mathrm{tr}(\vec{B}_l^{-1}\Sigmab_l),~l=1,\ldots,C,
\end{array}
\end{align}
where   ${\footnotesize\vec{C}^{(k_{\mathrm{c}})}=\dfrac{1}{2}\big( \vec{Z}^{(k_{\mathrm{c}}+1)}+\vec{V}^{(k_{\mathrm{c}}+1)}-\displaystyle\dfrac{\vecgreek{\Lambda}_{\mathrm{z}}^{(k_{\mathrm{c}})}+\vecgreek{\Lambda}_{\mathrm{v}}^{(k_{\mathrm{c}})}}{\rho}\big)}$ and \\
${\alpha_i^{(k_{\mathrm{c}})}= \big(\beta_1 q_i^{(k_{\mathrm{c}})}-\beta_2 \log|\Sigmab_l^{(k_{\mathrm{c}})}|q_i^{(k_{\mathrm{c}})}\big)}$.

\subsection{Inner Loop}
After running the outer loop for some pre-defined $K_{\mathrm{c}}$ iterations, we detect the set of the estimated active clusters $\mathcal{\hat{S}}=\bigcup_{j \in \mathcal{J}} \mathcal{C}_j$, where $l \in \mathcal{J}$ if there exists $i \in \mathcal{C}_l$ such that $\|\vec{x}_i\|>\epsilon$, where $\epsilon>0$ is a small predefined parameter.
In the inner loop, the proposed algorithm aims to detect  the  active UEs belonging to $\mathcal{\hat{S}}$  by using the separable sparsity-promoting prior
${-\log  p(\vec{X}) \propto  \sum_{i \in \mathcal{\hat{S}}}\log(\|\vec{x}_i\|+\epsilon_0)}$. Furthermore, we apply the MM approximation to linearize the concave   $-\log  p(\vec{X})\approx \sum_{i \in \mathcal{\hat{S}}} g_i^{(k_{\mathrm{c}})}\|\vec{x}_i\|,$
where $k_\mathrm{u}$ is inner loop iteration index and 
${g_i^{(k_\mathrm{u})}=\big( \|\vec{x}_i^{(k_\mathrm{u})}\|+\epsilon_0\big)^{-1}}$.
The optimization problem for the inner loop is given by
\begin{equation}\footnotesize
\begin{array}{ll}
      &\hspace{-4mm}\{\hat{\vec{X}}_\mathcal{\hat{S}}^{(k_\mathrm{u}+1)},\Sigmab^{(k_\mathrm{u}+1)}\} =\displaystyle\underset{\vec{X}_\mathcal{\hat{S}},\Sigmab}{\min}~\displaystyle\frac{1}{2}\|\vec{Y}-\vec{\Phi}_\mathcal{\hat{S}}\vec{X}_\mathcal{\hat{S}}\tran\|_{\mathrm{F}}^{2}+\beta_1 \sum_{i \in \mathcal{\hat{S}}} g_i^{(k_\mathrm{u})} \|\vec{x}_i\| \\
      &\hspace{-4mm}\displaystyle+\beta_2\sum_{l\in \mathcal{J}}\sum_{i \in \mathcal{\hat{S}}_l}\vec{x}_{i}\herm\Sigmab_l\vec{x}_{i}-\beta_2\sum_{l\in \mathcal{J}}  \log |\Sigmab_l| \big(\sum_{i \in \mathcal{C}_l} g_i^{(k_\mathrm{u})}\|\vec{x}_i\|+\beta_3Ld\big) \\
      &\hspace{-4mm}\displaystyle +\beta_3 L\sum_{l\in \mathcal{J}} \mathrm{tr}(\vec{B}_l^{-1}\Sigmab_l),
\end{array}\label{map_S_inner}
\end{equation}
where $\vec{\Phi}_\mathcal{\hat{S}}$ and $\vec{X}_\mathcal{\hat{S}}$ denote the matrices $\vecgreek{\Phi}$ and $\vec{X}$, respectively, restricted to the set of  the estimated active clusters, $\mathcal{\hat{S}}$. 


\subsection{ADMM Update Rules}
Owing to the applied splitting techniques, all the sub-problems \eqref{eq::z(k_c+1)_sec2}--\eqref{eq::sigma(k_c+1)_sec} are convex, thus, can be solved analytically via closed-form formulas. Further, the optimization over $\vec{X}$, $\vec{V}$, and $\Sigmab$ is separable over the UEs and the clusters, allowing for parallel updates. Thus, the outer loop update is given as
\begin{equation}\footnotesize\label{ADMM_update}
    \begin{cases}
       \vec{Z}^{(k_{\mathrm{c}}+1)}=  \big(\rho \vec{X}^{(k_{\mathrm{c}})} +\vec{\Lambda}^{(k_{\mathrm{c}})}+\vec{Y}\tran\vec{\Phi}^*\big) \big( \vec{\Phi}\tran \vec{\Phi}^*+\rho \vec{I}_N\big)^{-1},\\
       \vec{v}_i^{(k_{\mathrm{c}}+1)}=   \big(\beta_2\Sigmab_i^{(k_{\mathrm{c}})}+\rho\vec{I}_M \big)^{-1}(\rho \vec{x}_i^{(k)}+\vecgreek{\lambda}_{\mathrm{v}i}^{(k_{\mathrm{c}})}),\quad \forall i \in  \mathcal{N}, \\
      \vec{x}_i^{(k_{\mathrm{c}}+1)} = \max{\big\{0,\norm{\vec{c}_i^{(k_{\mathrm{c}})}}-\frac{\alpha_i^{(k_{\mathrm{c}})}}{2\rho}\big\}}\norm{\vec{c}_i^{(k_{\mathrm{c}})}}\vec{c}_i^{(k)},\quad\forall i \in \mathcal{N},\\
      \Sigmab_l^{(k_{\mathrm{c}})+1}=\big(\beta_2\sum_{i \in \mathcal{C}_l}\vec{v}^{(k_{\mathrm{c}})}_{i}\vec{v}_{i}^{(k_{\mathrm{c}})\herm}+\beta_3L\vec{B}_l^{-1}\big)\mu^{(k_{\mathrm{c}})}, l=1,\ldots,C.
    \end{cases}       
\end{equation}
The update rules for the inner loop, i.e., the solution to \eqref{map_S_inner} is similar to \eqref{ADMM_update} but with changing $q_i^{(k_{\mathrm{c}})}$ to $g_i^{(k_\mathrm{u})}$. Algorithm 1 provides the proposed ADMM solution  to  optimization problem \eqref{map_S}.

\begin{algorithm}[t]
\DontPrintSemicolon
   \footnotesize \KwInput{$\vec{\Phi}$, $\{\vec{B}_l\}_{l=1}^C$,
   $\scriptsize\beta_1$,$\beta_2$, $\beta_3$ ,$\rho$, $\epsilon_0$, $\epsilon$,   $k_{u_{\mathrm{max}}}$, $k_{c_{\mathrm{max}}}$, $\scriptsize K_{\mathrm{c}}$.}
\footnotesize \Kwinitialize{$\scriptsize \vec{X}^{(0)},\vec{V}^{(0)},\vec{Z}^{(0)},\vecgreek{\Lambda}_{\mathrm{v}}^{(0)},\vecgreek{\Lambda}_{\mathrm{z}}^{(0)}, {k_\mathrm{u}=1, k_{\mathrm{c}}=1.}$}
\footnotesize BS receives $\vec{Y}$, compute and store $\big( \vec{\Phi}\tran \vec{\Phi}^*+\rho \vec{I}_N\big)^{-1}$ 
 
   \While{$k_{\mathrm{c}}<k_{c_{\mathrm{max}}}$ $\mathrm{or}$ $\| \vec{X}^{(k_{\mathrm{c}})}-\vec{X}^{(k-1)} \|<\epsilon$ }
   {

  $\tiny\vec{Z}^{(k_{\mathrm{c}}+1)}=  (\rho \vec{X}^{(k_{\mathrm{c}})} +\vec{\Lambda}^{(k_{\mathrm{c}})}+\vec{Y}\tran\vec{\Phi}^*)( \vec{\Phi}\tran \vec{\Phi}^*+\rho \vec{I}_N)$ 
 \; 
  $\scriptsize\vec{v}_i^{(k_{\mathrm{c}}+1)}=  \big(\beta_2\Sigmab_i^{(k_{\mathrm{c}})}+\rho\vec{I}_M \big)^{-1}(\rho \vec{x}_i^{(k)}+\vecgreek{\lambda}_{\mathrm{v}i}^{(k_{\mathrm{c}})}), {i=1,\ldots,N}$ 
 \;\vspace{-.3cm}
 $\scriptsize\vec{x}_i^{(k_{\mathrm{c}}+1)}= \frac{\max{\big\{0,\norm{\vec{c}_i^{(k_{\mathrm{c}})}}-\frac{\alpha_i^{(k_{\mathrm{c}})}}{2\rho}\big\}}\vec{c}_i^{(k_{\mathrm{c}})}}{ \norm{\vec{c}_i^{(k_{\mathrm{c}})}}}, {i=1,\ldots,N}$
 \;
  $ \scriptsize\vec{\Sigmab}_l^{(k_{\mathrm{c}})+1}=\big(\beta_2\sum_{i \in \mathcal{C}_l}\vec{v}_{i}\vec{v}_{i}\herm+\beta_3L\vec{B}_l^{-1}\big)\mu^{(k_{\mathrm{c}})},l=1,\ldots,C$
 \;\vspace{-.3cm}
 $\scriptsize\vecgreek{\Lambda}_{\mathrm{z}}^{(k_{\mathrm{c}}+1)}=  \vecgreek{\Lambda}_{\mathrm{z}}^{(k_{\mathrm{c}})}+\rho\big(   \vec{X}^{(k_{\mathrm{c}}+1)}-\vec{Z}^{(k_{\mathrm{c}}+1)} \big)$
 \;
 $\scriptsize\vecgreek{\Lambda}_{\mathrm{v}}^{(k_{\mathrm{c}}+1)}=  \vecgreek{\Lambda}_{\mathrm{v}}^{(k_{\mathrm{c}})}+\rho\big(   \vec{X}^{(k_{\mathrm{c}}+1)}-\vec{V}^{(k_{\mathrm{c}}+1)} \big)$
 \;

\uIf{\big($k_{\mathrm{c}}\ \mathrm{mod}\ K_{\mathrm{c}}\big) = 0$}{
 $\scriptstyle\mathcal{\hat{S}}=\bigcup_{j \in \mathcal{J}} \mathcal{C}_j$, $\{l \in \mathcal{J}: \exists i \in \mathcal{C}_l| \|\vec{x}_i\|>\epsilon\}$
 \;
  \While{\scriptsize$k_\mathrm{u}<k_{\mathrm{u}_{{\mathrm{max}}}}$ }
   {
 Solve \eqref{map_S_inner} using the similar update rules as \eqref{ADMM_update}, but using $\scriptsize g_i^{(k_\mathrm{u})}=\big(\|\vec{x}_i^{(k_\mathrm{u})}+\epsilon_0\|\big)^{-1}$\;
$k_{\mathrm{u}}\leftarrow{k_{\mathrm{u}}+1}$\;

 }

$\scriptstyle\vec{X}_{\mathcal{\hat{S}}}^{(k_{\mathrm{c}})}=\vec{X}_{\mathcal{\hat{S}}}^{(k_\mathrm{u})}$, $\scriptstyle\vec{Z}_{\mathcal{\hat{S}}}^{(k_{\mathrm{c}})}=\vec{Z}_{\mathcal{\hat{S}}}^{(k_\mathrm{u})}$, $\scriptstyle\vec{V}_{\mathcal{\hat{S}}}^{(k_{\mathrm{c}})}=\vec{V}_{\mathcal{\hat{S}}}^{(k_\mathrm{u})}$, $\scriptstyle\vecgreek{\Lambda}_{\mathrm{z}_{\mathcal{\hat{S}}}}^{(k_{\mathrm{c}})}=\vecgreek{\Lambda}_{\mathrm{z}_{\mathcal{\hat{S}}}}^{(k_{\mathrm{u}})}$,  $\scriptstyle\vecgreek{\Lambda}_{\mathrm{v}_{\mathcal{\hat{S}}}}^{(k_{\mathrm{c}})}=\vecgreek{\Lambda}_{\mathrm{v}_{\mathcal{\hat{S}}}}^{(k_{\mathrm{u}})}$\;
  $k_\mathrm{u}=1$,\;
 }
$\scriptsize k_{\mathrm{c}}\leftarrow{k_{\mathrm{c}}+1}$\;
}
\caption{The proposed JUICE algorithm}
\end{algorithm}

\section{Numerical Results}
Consider a network with one BS equipped with ${M=20}$ antennas serving a  ${N=500}$ UEs distributed equally over ${C=20}$ clusters, where  ${K=16}$  UEs are active at each  $T_\mathrm{c}$.  Each UE is assigned a  unique unit-norm pilot sequence that is generated from an i.i.d. complex Bernoulli distribution. 
We set each $\vec{B}_l,  \forall l $, as $\vec{B}_l=\zeta \Psib_l+(1-\zeta) \frac{1}{L}\sum_{i\in \mathcal{C}_l}\Sigmab_i$, where $\Psib_l$ is a random semi-definite Hermitian matrix to model the error in the prior knowledge on the true precision matrix $\Sigmab_l$, whereas the parameter $\zeta$ controls the level of average  mismatch between $\Sigmab_i$ and $\vec{B}_l$, $\forall i \in \mathcal{C}_l$, and it is set as $\zeta=0.1$.


Channel estimation  is quantified in terms of normalized mean square error (NMSE) defined as  $\frac{\mathbb{E}\left [\| \vec{X}-\hat{\vec{X}}_\mathcal{S}\|_{\mathrm{F}}^2 \right ]}{\mathbb{E}\left[\| \vec{X}\|_{\mathrm{F}}^2 \right]},$
where the expectation is computed via Monte-Carlo averaging over all sources of randomness. UEs activity detection  is quantified in terms of  support recovery  rate (SRR), defined as $\frac{\vert \mathcal{S} \cap \hat{\mathcal{S}}\vert}{\vert \mathcal{S} - \hat{\mathcal{S}}\vert+K}$, where $\hat{\mathcal{S}}$ denotes the detected support by a given algorithm.

\begin{figure}[t!]
   \begin{subfigure}[b]{0.22\textwidth}    \includegraphics[width=\linewidth]{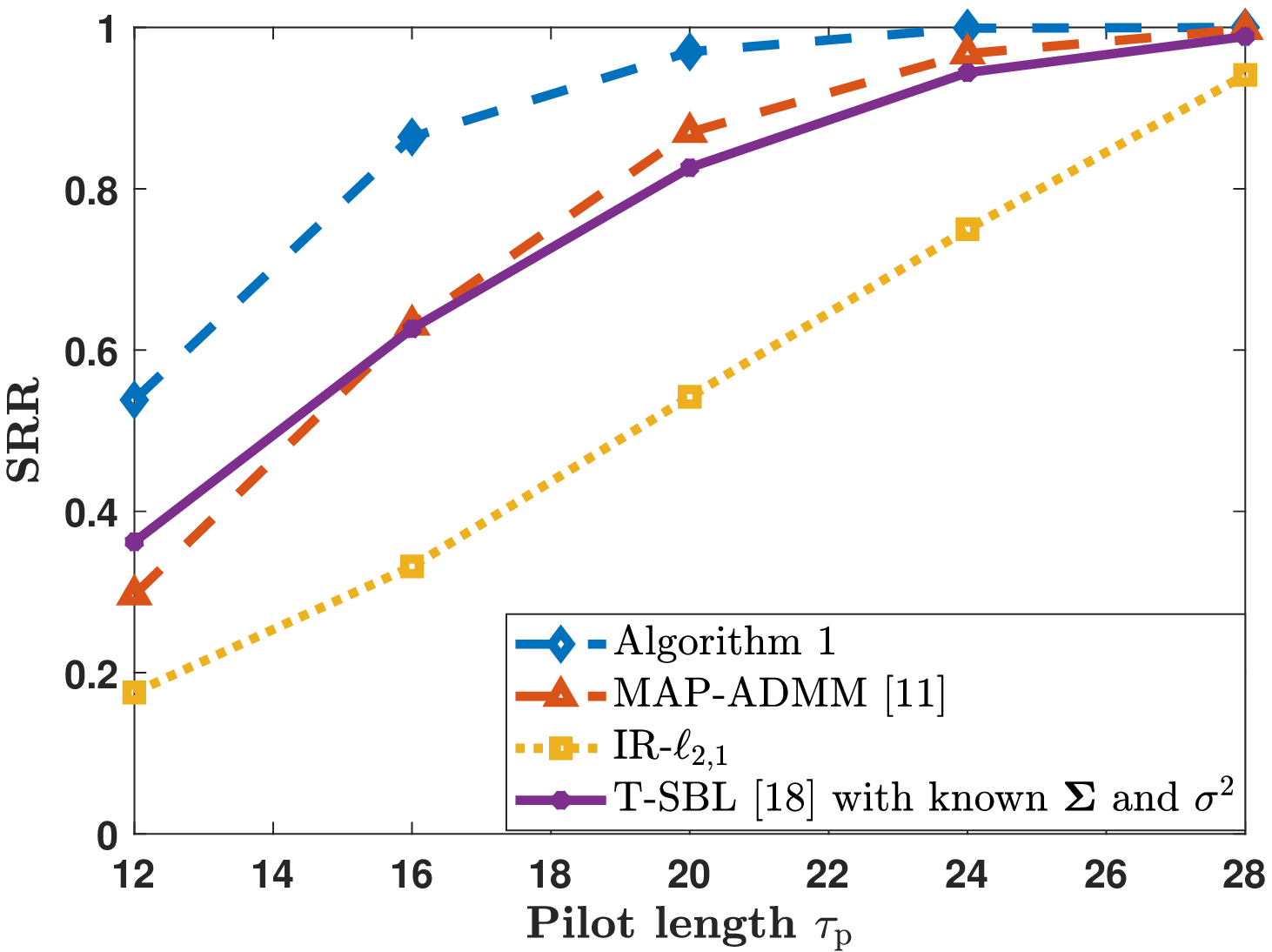}  \caption{}
    \label{mse_clus}
\end{subfigure}
   \hfill
    \begin{subfigure}[b]{0.22\textwidth}
  \includegraphics[width=\linewidth]{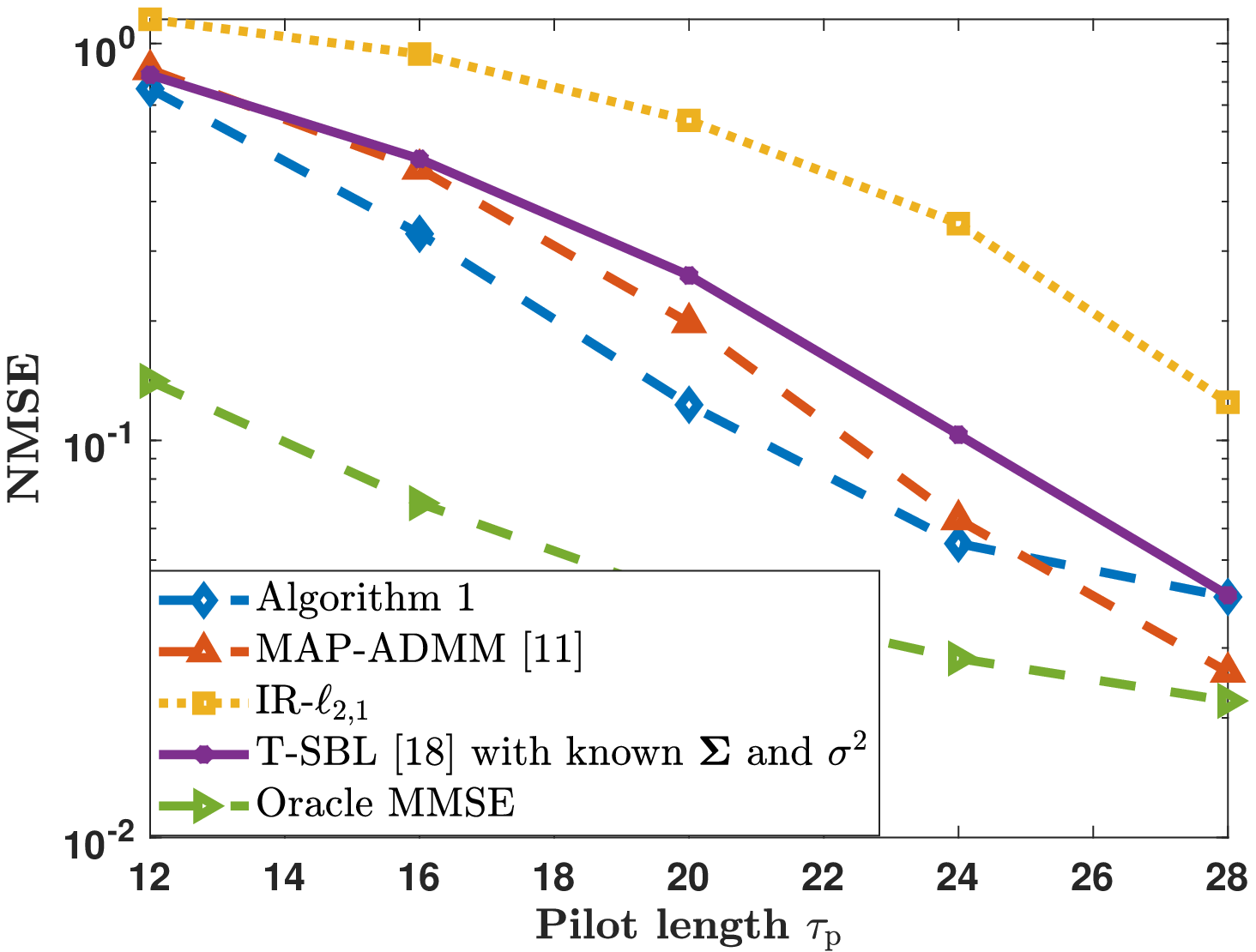}   \caption{}
     \label{mse_clus}
\end{subfigure}\vspace{-3mm}
\caption{Performance evaluation with  clustered activity pattern. }
\label{fig:results_cluster}
\vspace{-3mm}
\end{figure}

\begin{figure}[t!]
\begin{subfigure}[b]{0.22\textwidth}
    \includegraphics[width=\linewidth]{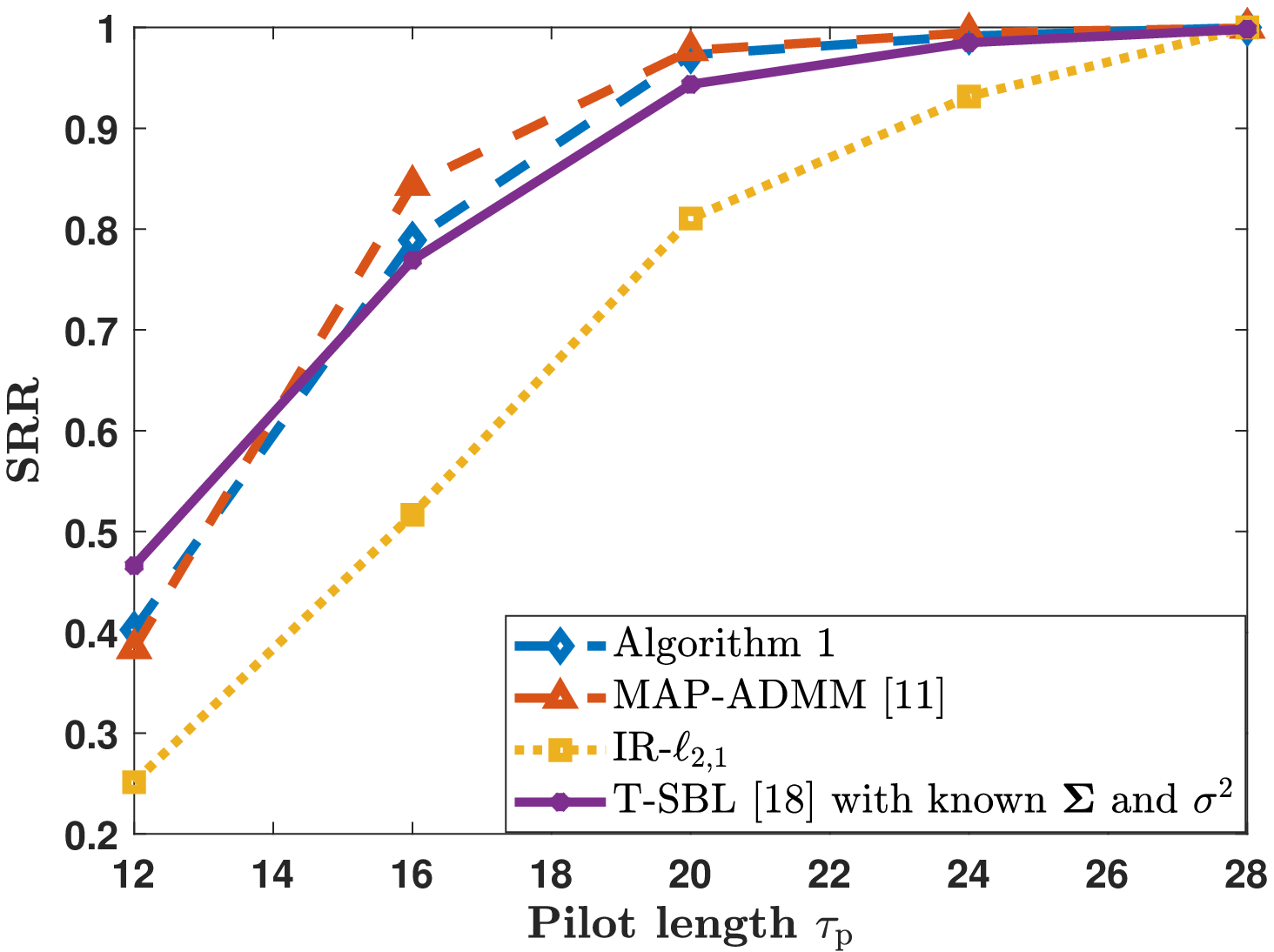}
    \caption{}
    \label{fig:srr_rand}
\end{subfigure}
   \hfill
    \begin{subfigure}[b]{0.22\textwidth}
  \includegraphics[width=\linewidth]{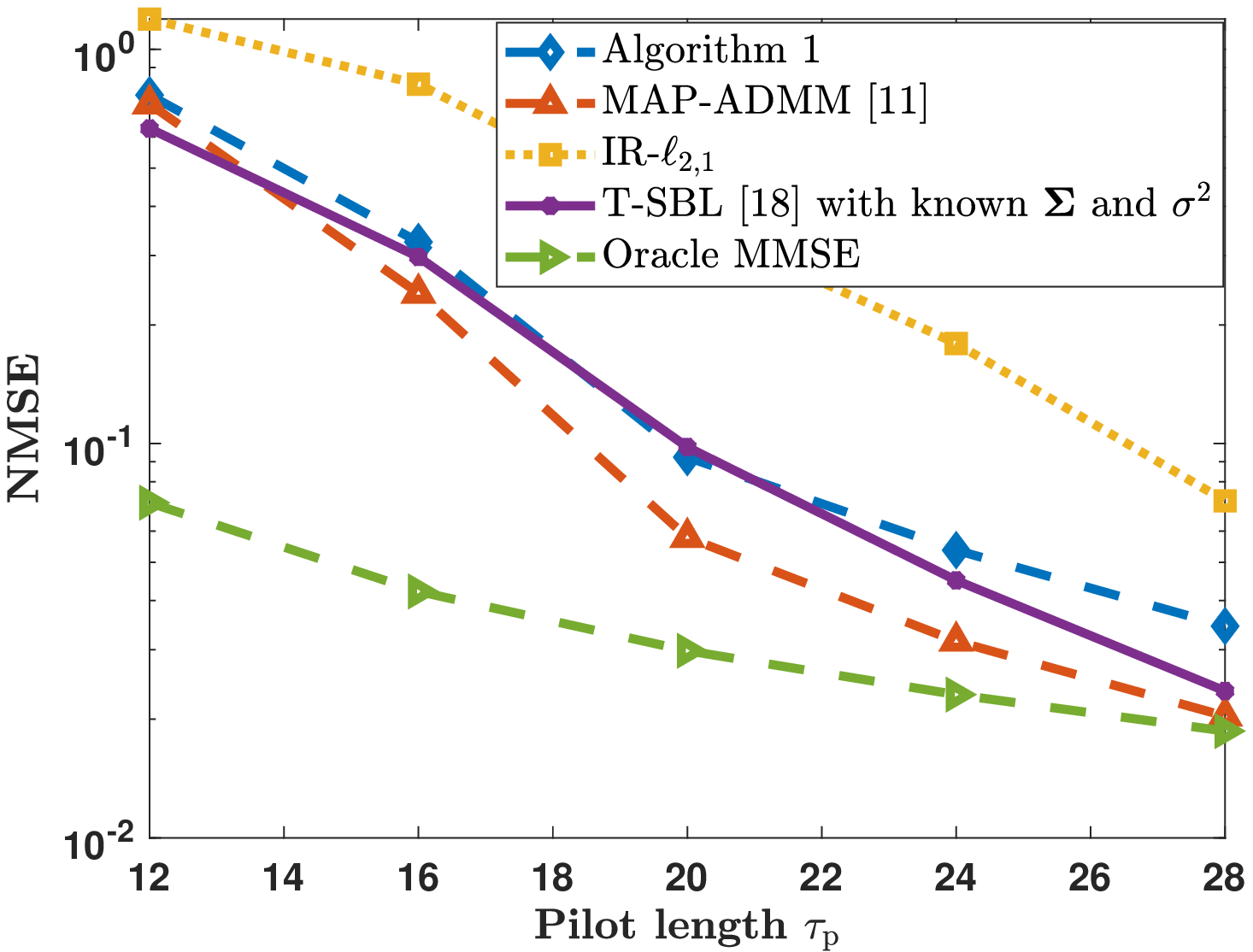}
     \caption{}
     \label{fig:mse_rand}
\end{subfigure}\vspace{-3mm}
\caption{ Performance evaluation with  random activity pattern. }
\label{fig:results_rand}
\vspace{-6mm}
\end{figure}


The performance of proposed Algorithm 1 is compared with: 1) iterative reweighted $\ell_{2,1}$ (IR-$\ell_{2,1}$) via ADMM  \cite{Djelouat-Leinonen-Juntti-21-icassp}, 2) MAP-ADMM \cite{djelouat2021spatial}, and  3) T-SBL \cite{zhang2011sparse}, where for  MAP-ADMM and T-SBL,  both $\Sigmab$ and $\sigma^2$ are known to the BS.     For an optimal NMSE benchmark, we consider the oracle minimum mean square error (MMSE) estimator having prior  knowledge on the true set of active UEs and their covariance matrices.




We consider two cases for UEs' activity pattern: 1) clustered: with 2 active clusters, each with $8$ active UEs, and
2) random, where ${K=16}$ active UEs are randomly distributed among the clusters.


Fig.\ \ref{fig:results_cluster}(a)  presents the  SRR  against the pilot length for the clustered activity pattern. 
The result shows that Algorithm 1 provides the highest SSR, which can be explained by the fact that Algorithm 1 effectively exploits the clustered structure of the activity pattern using the proposed prior functions.
Similarly, the proposed algorithm moderately outperforms MAP-ADMM and T-SBL in terms of NMSE (Fig. \ \ref{fig:results_cluster}(b)). In summary, Fig. \ref{fig:results_cluster} clearly shows the utility of the proposed algorithm in JUICE with clustered activity patterns.

Fig. \ref{fig:results_rand}(a) illustrates the SRR for the random activation scenario. Algorithm 1 achieves performance similar to MAP-ADMM and T-SBL while providing significant gains over IR-$\ell_{2,1}$. On the other hand, Fig.  \ref{fig:results_rand}(b) shows that while the proposed algorithm clearly outperforms  IR-$\ell_{2,1}$ in terms of channel estimation accuracy, it provides just a slightly inferior NMSE performance compared to T-SBL and MAP-ADMM. These results show the robustness of the proposed algorithm to the structure of the activity pattern: the pattern can deviate from the desired clustered form to fully random activity, rendering the algorithm effective for a wide range of UE activity models.


\vspace{-.2cm}
\section{Conclusion}
We derived a Bayesian framework to address the JUICE with clustered activity in mMTC under correlated MIMO channels without prior knowledge of the exact CDI. To encourage solutions with cluster sparsity, we proposed  a cluster-sparsity-promoting function that correlates the activity of the UEs   belonging to the same cluster. We developed an ADMM algorithm that provides a computationally-efficient solution via a sequence of closed-form update rules. The numerical results showed the improvement and robustness brought by the proposed algorithm.


\bibliographystyle{IEEEbib}
\mybibliography
\end{document}